%% file: main.tex
\documentclass[]{spie}  

\usepackage{amsmath,amsfonts,amssymb}
\usepackage{graphicx}
\usepackage[colorlinks=true, allcolors=blue]{hyperref}

\input{journaux.tex}

\title{ESCAPE project: investigating active observing strategies and post-processing methods for exoplanet high-contrast imaging with future space missions}

\author[a]{Elodie Choquet}
\author[a]{Lisa Altinier}
\author[a]{Nicolás Godoy}
\author[a]{Alexis Lau}
\author[a]{Arthur Vigan}
\author[b]{David Mary}
\affil[a]{Aix Marseille Univ, CNRS, CNES, LAM, Marseille, France}
\affil[b]{Université Côte d’Azur, Observatoire de la Côte d’Azur, CNRS, Laboratoire Lagrange, Bd de l’Observatoire, CS 34229,
06304 Nice Cedex 4, France}

\authorinfo{Contact author: elodie.choquet@lam.fr}

\begin{document} 
\maketitle

\begin{abstract}
 The search for biosignatures in potentially habitable exoplanets is one of the major astrophysics’ drivers for the coming decades, and the prime science goal of the HWO NASA mission, a large UV-Optical-IR space telescope to be launched in the 2040s. To reach this goal, it will be equipped with state-of-the-art high-contrast spectro-imaging capabilities enabling the detection of exoplanets $10^{10}$ times fainter than their host stars, a formidable challenge given today's best detection limits at $\sim10^{-6}$ contrast levels. 
 This goal puts stringent constraints on the entire observatory, and demands the optimization at the system level to leverage the performance of individual sub-systems. However, while image processing techniques are a key asset to reach the ultimate performance, the science and technological definition of the mission concepts mostly rely on  the coronagraph and wavefront control to reject the starlight, assuming a conservative gain of $\sim10$ in sensitivity from image processing, extrapolated from performance obtained with classical techniques on Hubble observations. 
In the ESCAPE project, we investigate integrated solutions for optimizing the observing methods and data processing techniques with future space telescopes, making use of their wavefront sensors and deformable mirrors. The Roman Space Telescope, scheduled for launch in 2026, will be a critical milestone to demonstrate key technologies ahead of HWO with the Coronagraph instrument, and is thus a unique opportunity to also test and validate innovative image processing techniques. Here we present the rational, methodology, and timeline of the ESCAPE project.
\end{abstract}


\section{INTRODUCTION} \label{sec:intro} 

Determining the frequency of life in the Universe is one of the greatest challenges of the next decades. While the frequency of Earth-like planets has been constrained by exoplanet transit detection missions\cite{Kopparapu2018,Bryson2021}, determining the diversity of their atmospheric composition, the frequency of habitable conditions, and the existence of exo-biosignatures entails the spectral characterization of dozens of temperate rocky planets (``exoEarths''). This is best achieved with direct observations of the exoplanet itself in order to reach the necessary signal-to-noise ratio (S/N) in a reasonable observing time. To this goal, following the US National Academies' recommendation from the Astronomy and Astrophysics 2020 (Astro2020) Decadal Survey\cite{Astro2020}, NASA has prioritized a large UV-optical-IR telescope, the Habitable World Observatory (HWO), as its next flagship astrophysics mission after the Roman Space Telescope, to be launched in the 2040s.  This observatory will be equipped with state-of-the-art high-contrast spectro-imaging capabilities, with as prime science goal to directly image a minimum sample of 25 exoEarths and study their atmosphere in reflected starlight.


Reaching this goal implies being able to distinguish sources $10^{-10}$ times fainter than their host star\cite{Traub2010} at only 50--100~mas separation for an exoEarth at 10--20~pc distance, i.e. $\sim3-6$ resolution elements ($\lambda/D$) away from the star center image with a 6m-diameter telescope at $\lambda=500$~nm. This represent a tremendous technical challenge, 5 orders of magnitude better than the typical detection limits of current state-of-the-art high-contrast imagers (e.g. contrast limits of $1-5\times 10^{-5}$ at $5~\lambda/D$ separation with the SPHERE-IFS\cite{Beuzit2019}, GPI\cite{Macintosh2014}, and JWST-NIRCam instruments\cite{Carter2023}). 
To reach this goal, HWO will be equipped with state of the art technologies for high-contrast imaging, that are currently investigated and will be further developed in the coming years as part of the Great Observatory Maturation Program (GOMAP). Critical aspects include telescope architectures, coronagraph designs, deformable mirror technologies, low order wavefront sensing and control, focal plane high-order wavefront sensing and dark hole algorithms, and photon-counting detectors. These technologies must also be optimized altogether at the system level, so that ultimately, after propagation from the telescope primary mirror down to the science focal plane, the optical wavefront is stabilized down to $\sim$10~pm rms, an extremely constraining requirement necessary to reach $10^{-10}$ raw contrast limits\cite{Leboulleux2018,Laginja2021,Pueyo2022}.

In this technology roadmap, developing and optimizing image post-processing techniques can be a key asset. Last pillar of the high-contrast imaging process, post-processing makes use of all relevant existing data to  disentangle \textit{a posteriori} the planet signal from the remaining stellar contamination in the science data, effectively pushing exoplanet detection limits beyond the instrument raw contrast limits.  Post-processing algorithms are strongly aided by clever observing strategies, that must be also optimized to acquire the best flight data to calibrate the residual stellar contamination in the science datasets at the post-processing stage. Post-processing is an active field of research both for ground-based and space high-contrast imaging observations (e.g. ref. \citenum{Vogt2011,Singh2016,Soummer2014a,Lajoie2016,Potier2022}), yet the science and technological definition studies for the HWO precursor concepts (LUVOIR, HabEx) provisioned a conservative gain of 10x from post-processing in the contrast limit budget, expected from classical techniques optimized for the passive Hubble and James Webb space telescopes\cite{Stark2019,Stark2024,LUVOIRTeam2019,Gaudi2020} (Classical RDI or ADI). Developing better post-processing methods, with observing strategies optimized for specifics of HWO observations, has the capacity to better model and remove the fluctuating residual starlight and improve its detection limits. This could in turn either improve the mission exoplanet yield or relax its stringent picometer-level stability constraints.

In the meantime, the next NASA astrophysics mission, the Nancy Grace Roman Space Telescope (hereafter Roman), planned for launch late 2026 to early 2027, includes the Coronagraph instrument, which will demonstrate high-contrast imaging operations with a number of state-of-the art technologies\cite{Bailey2023}. Specifically, the Roman Coronagraph is equipped with nanometer surface error super-polished mirrors\cite{Roulet2020}, optimized coronagraphic masks\cite{Riggs2021}, a Zernike low-order wavefront sensor (LOWFS)\cite{Shi2018}, two $48\times48$ actuators deformable mirrors (DMs), and a photon-counting EMCCD detector, all key technologies for the HWO design studies. In particular, the Roman Coronagraph will demonstrate high-order wavefront control for the first time in space thanks to its two DMs, using pairwise probing and electric field conjugation\cite{Giveon2007a,Giveon2007b} (EFC) to create a ``dark zone'' optimized for very high-contrast limits at short separations from the star. It is tasked to demonstrate $10^{-7}$ contrast limits at 5$\sigma$ in a 6--9$\lambda/D$ area from a star in a broad 10\% bandwidth centered at $\lambda=575$~nm (threshold technical requirement, ``TTR5''), but is in fact designed to reach better than $10^{-8}$ contrast limits in the same bandpass and within 3 to 9$\lambda/D$ from the star\cite{Krist2023}. These unprecedented detection limits will enable for the first time the visible reflected-light atmospheric characterization of giant exoplanets orbiting main sequence stars on similar spatial scales as the solar system (5--10~au)\cite{Bailey2018}. As such, the Roman Coronagraph is also a unique opportunity to test and demonstrate innovative observing strategies and post-processing methods adapted to such unexplored contrast levels and state-of-the art technologies, ahead of the definition of the HWO mission.

The ESCAPE project (Exoplanetary Systems with a Coronagraphic Active Processing Engine) is an ERC program funded from 2022 to 2027 by the European Union to develop observing strategies and post-processing techniques optimized for the future exoplanet high-contrast imagers in space, improve their detection limits, and allow for the detection and characterization of a larger sample of nearby exoplanets. In particular, it will explore how the active components, wavefront sensors and deformable mirrors, new to these future space missions, can be used to optimize the detection limits at post-processing. The goal of the project is to develop these methods in time for operations with the Roman Coronagraph, and demonstrate them on-sky during its allocated 90 days of science observations, to serve for future developments for the HWO mission. In this paper we provide an overview of the ESCAPE project, detailing its methodology and current status in Sec.~\ref{sec:methodo}.

\section{OVERVIEW OF THE ESCAPE PROGRAM}\label{sec:methodo}

\subsection{Rational and Objectives}
While being an active field of research, post-processing methods are dissimilarly developed for ground-based vs. space exoplanet imagers. While both type of facilities started by using classical Reference star Differential Imaging (RDI) as a starlight subtraction method\cite{Smith1984,Schneider1999}, ground-based observations quickly develop dedicated observing strategies (short exposures, ADI\cite{Marois2006}, SDI\cite{Sparks2002}) and advanced starlight modeling algorithms (e.g. LOCI\cite{Lafreniere2007}) in order to mitigate the speckle noise resulting from first generation of adaptive optics systems\cite{Racine1999,Soummer2007,Hinkley2007}. The advent of second-generation extreme adaptive optics systems, with dedicated high-order wavefront sensors and deformable mirrors has triggered interesting new post-processing methods that make use of the unprecedented wavefront and image stability (e.g. \citenum{Flasseur2018}). In particular new methods using the deformable mirrors and sensor telemetry to estimate a posteriori the stellar intensity distribution or the coherent electric field and subtract it from the science data have emerged (e.g. LOWFS telemetry + focal plane image dictionaries\cite{Vogt2011,Sing2016}) and, for some, been tested on sky (e.g. CDI\cite{Potier2022}). Still, on ground-based telescopes, these methods remain limited to a few $10^{-6}$ contrast limits at best within $15~\lambda/D$\cite{Potier2022}, several orders of magnitude above the contrast regimes of the Roman Coronagraph and HWO.

In space, classical post-processing techniques (RDI and roll subtraction\cite{Lowrance2005}) have remained the only starlight subtraction methods for HST coronagraphic observations until quite recently. The development of archival reference star libraries\cite{Lafreniere2009,Soummer2011,Hagan2018,Ren2017} ultimately significantly improved the detection limits, as they provide enough PSF realizations to sample the wavefront diversity of the instrument over its entire lifetime and allow to optimize the starlight subtraction in any science dataset at any point in time with the help of advanced post-processing algorithm (e.g. PCA\cite{Soummer2012}). This method, while being particularly successful (see e.g. \citenum{Soummer2014,Choquet2016,Choquet2017,Choquet2018,Marshall2018,Marshall2023}), is relatively inefficient as it requires to accumulate reference star observations for many cycles/years to finally provide enough image diversity to achieve a significant gain over classical subtraction techniques that use contemporaneous reference star observations. A smarter observing strategy was later developed, the small grid dither method\cite{Soummer2014a,Rajan2015,Debes2019}, which uses the Fast Steering Mirror (FSM) to methodically sample the coronagraph response to tip-tilt misalignment during the reference star observations contemporaneous to the science target visits. This method proved to be particularly efficient for instruments where tip-tilt pointing errors are the dominating contrast limitation, and it doesn't require to wait for many observations to randomly sample the tip-tilt error parameter space. This observing method is now also the default for JWST coronagraphic observations\cite{Lajoie2016, Carter2023}.

\begin{figure} [ht]
   \begin{center}
   \begin{tabular}{c} 
   \includegraphics[height=8cm]{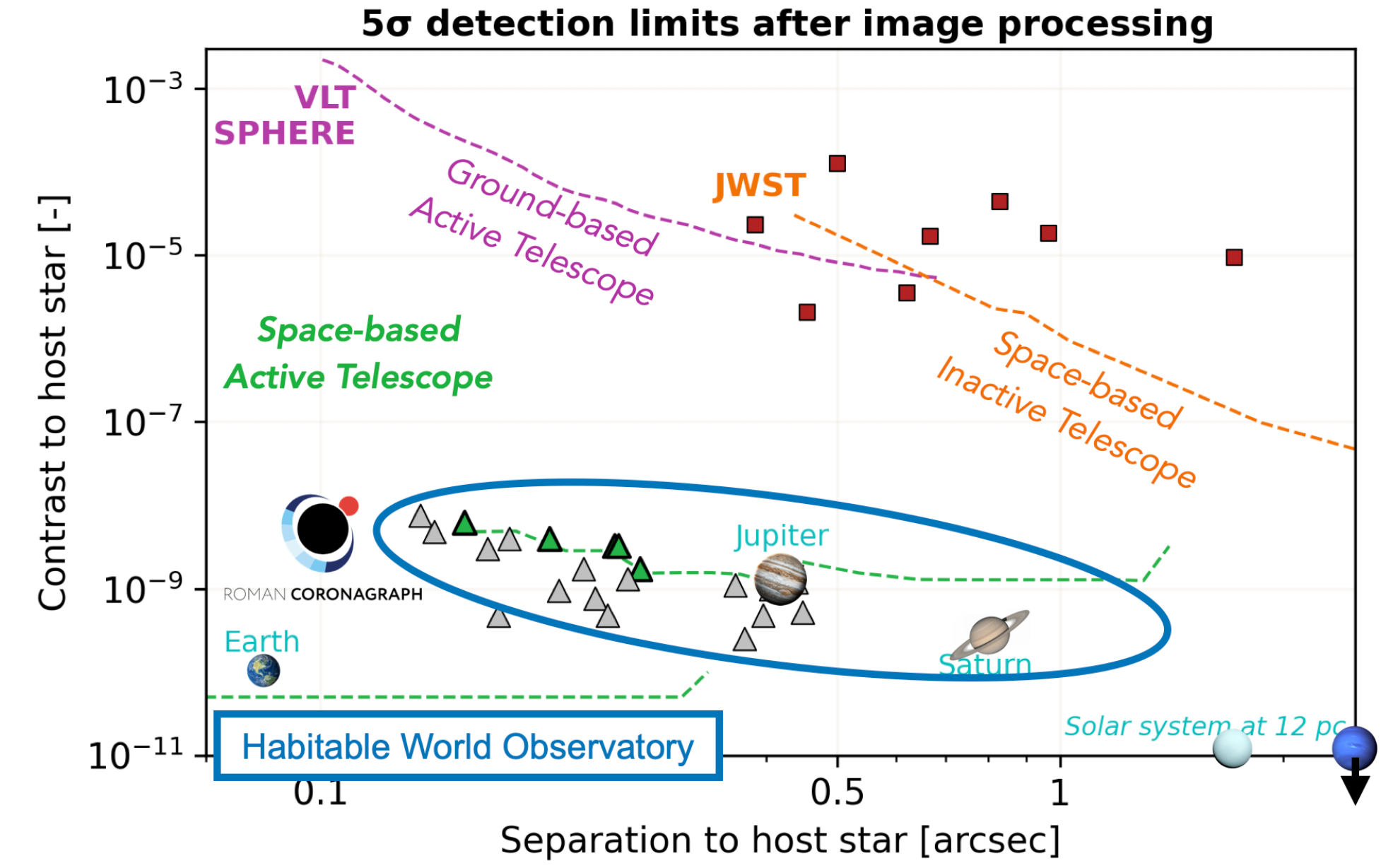}
   \end{tabular}
   \end{center}
   \caption{ \label{fig:contrastplot}Median detection limits of state-of-the-art instruments on the ground (SPHERE\cite{Beuzit2019}, purple line) and in space (JWST\cite{Beichman2010}, orange), and expected limits for the future active space instruments (Roman Coronagraph and HWO). The red square markers feature the self-luminous young giant planets discovered with  ground-based high-contrast imagers in the near-infrared. The triangular markers indicate RV-detected Jovian planets that are within the expected detection limits of the Roman coronagraph (in green) and those just below these limits (in gray). In blue we highlight the two objectives of the ESCAPE project: push the detection limits of the Roman coronagraph to detect and characterize the Jovian planets at its detection limits, and inform the design of the Habitable World Observatory with improved post-processing techniques tested on Roman. Solar system planets as seen at 12~pc are indicated for comparison. Adapted from \texttt{https://github.com/nasavbailey/DI-flux-ratio-plot}\cite{Bailey2018}.}
   \end{figure}

Having high-order deformable mirrors and wavefront sensors in space opens new prospects to build upon these previous works and to push observing strategies and post-processing techniques to $10^{-9}$ contrast limits and beyond. Yet, the baseline observing strategy and post-processing algorithms planed the Roman coronagraph are still based on classical or PCA RDI, with standard reference star observations interleaved between science target visits once the dark hole has been generated with EFC\cite{Bailey2018}. The goal of the ESCAPE program is to develop such ``active'' post-processing techniques and to implement them on the Roman coronagraph, in order to demonstrate a gain on sensitivity limits over the baseline classical techniques. Demonstrating such an improvement would then allow to observe and characterize a larger sample of Jovian exoplanets known from radial velocity (RV) surveys, that are otherwise just at the sensitivity limits of the expected performance of the Roman coronagraph (see Fig.~\ref{fig:contrastplot}). Furthermore, it would provide critical information to feed the technical and science definition of HWO, possibly contributing to relaxing some of its stability constraints.

\subsection{Methodology}
\begin{figure} [ht]
   \begin{center}
   \begin{tabular}{c} 
   \includegraphics[height=7cm]{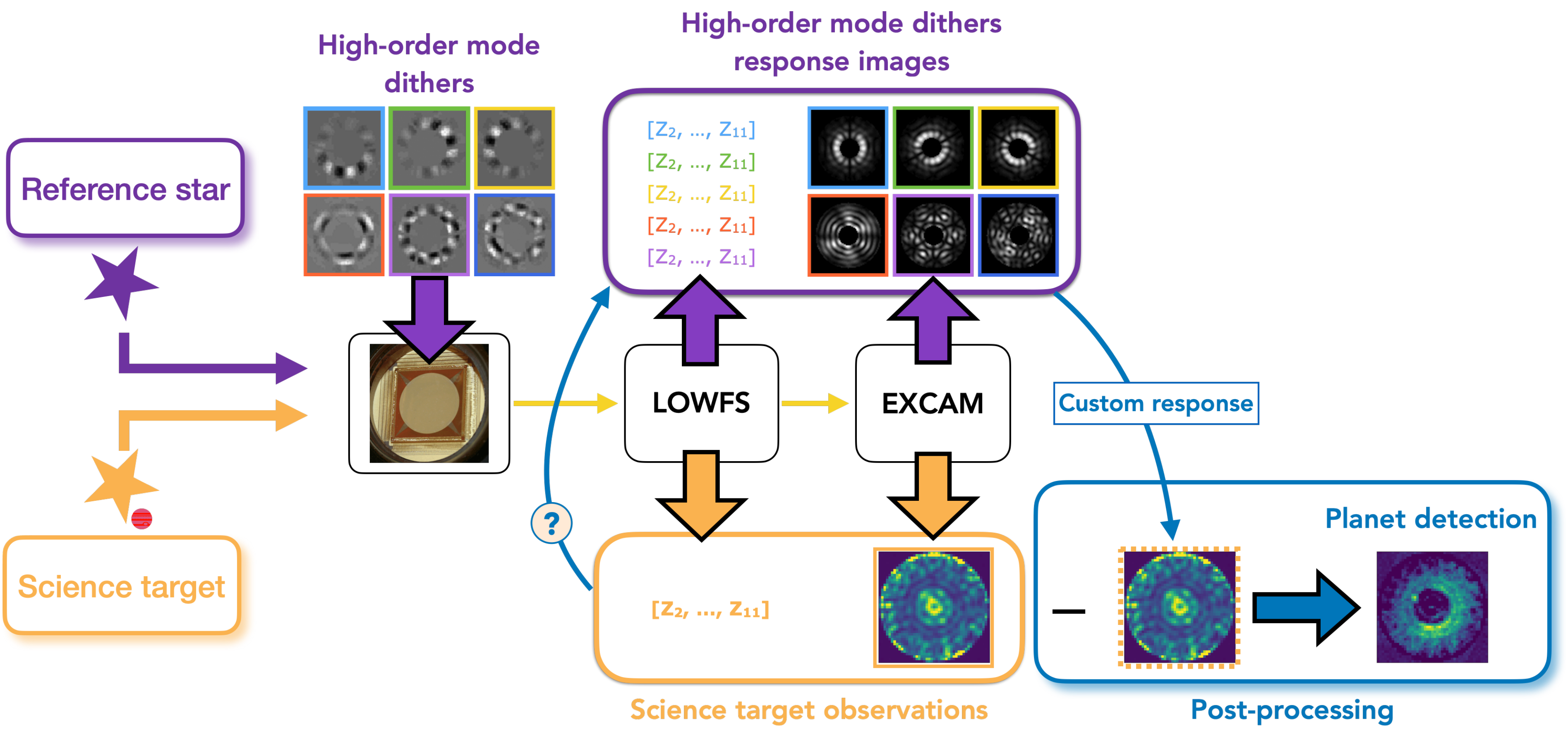}
   \end{tabular}
   \end{center}
   \caption{ \label{fig:solution} Proposed observing strategy and post-processing strategy for Roman Coronagraph operations. After the dark hole is created, high-order mode dithers are applied on the DM during the reference star observations and the corresponding science camera (EXCAM) images and low order wavefront sensor measurements are saved. After the science target observations, the science camera images are compared to the high-order mode dithered reference star image library using PCA-type algorithms to produce a custom speckle pattern that matches the science observations. This approach is a generalization of the tip-tilt and low-order mode dithers tested on JWST\cite{Soummer2014a} and Subaru\cite{Vogt2011,Singh2016} to high-order modes within a dark hole. In complement, the low-order wavefront sensor telemetry can be used to further inform the speckle pattern model by down-selecting or excluding specific images from the dither library depending on how they match with the science target telemetry. Simulated images and modes from \citenum{Krist2023,Girard2020}}
   \end{figure}

The methodology of the ESCAPE program to develop these active methods  follows two axes in parallel: 1/ use existing high-contrast imaging facilities (in particular JWST) to learn their limits with classical post-processing techniques and to push these limits using the telemetric measurements, and 2/ prepare for Roman coronagraph operations by developing observing strategies using its high-order deformable mirrors.
In Fig.~\ref{fig:solution}, we outline an integrated strategy combining both approaches that would be well adapted to Roman Coronagraph operations.  

\subsubsection{Learning from JWST and using telemetric measurements}
On the first axis, we are currently investigating the capabilities and limitations of the coronagraphic modes on JWST, space telescope that is not equipped with close-loop wavefront control but still shares the same extremely stable environment at L2 as Roman and HWO. In particular,  we are using GTO data (GTO 1241, PI M. Ressler) to learn the behavior, systematics, and science performance of MIRI coronagraphy on 4 science targets with known substellar companions probing different contrast regimes, while constraining their atmospheric and physical properties with the mid-infrared photometric measurements. Results on the first target of this program, HR 2562 B, will be soon published in Godoy et al. 2024a (accepted in A\&A). 

In parallel, we are investigating the fundamental limits of NIRCam coronagraphy, set by the photon noise, detector noise, and calibration noise sources, to compare them with the detection limits achieved after starlight subtraction with state-of-the-art post-processing techniques (small-grid dither strategy, RDI, PCA). Using public cycle 1 NIRCam coronagraphic programs (commissioning\cite{Girard2022}, ERS\cite{Carter2023}, and DDT\cite{Franson2024} data), Godoy et al. 2024b (these proceedings) used both an analytical and an MCMC-based noise propagation approach to estimated the theoretical noise limits in these datasets, and demonstrated a gap of a factor of 5 to 10 within 1'' separation between the measured post-processed detection limits and the photon noise limits. This indicates that the current PSF subtraction methods are not yet optimal and still leave important  starlight residuals at short separation, even at $10^{-6}$ contrast levels. This also opens prospects for improving these detection limits with better post-processing methods, already for JWST observations and without waiting for Roman coronagraph.

In particular, while not performing real-time wavefront control in closed loop, JWST still carries out regular wavefront measurements every 2--3 days through phase retrieval with the NIRCam instrument\cite{Perrin2016a,Rigby2023}. These telemetric measurements can readily be used to inform and improve image processing methods and to push detection limits. As part of ESCAPE we are now investigating how they can be used to create a hybrid PSF library, built with both empirical reference star images and WebbPSF-generated synthetic images, to improve post-processed detection limits. These development will also benefit to future Roman coronagraph observations, as low-order wavefront sensor and pairwise probe measurements will also be available a posteriori for the post-processing steps.

\subsubsection{Developing high-order mode dithering strategy for the Roman coronagraph}
In parallel, on the second axis, we are investigating observing strategies making use of the deformable mirrors to improve detection limits, in particular in the context of the Roman Coronagraph. Specifically, we are exploring how the tip-tilt dither strategy adopted on HST and JWST can be generalized to higher order aberrations using the Coronagraph DMs. While pointing errors are the main contrast limitation with the HST/BAR5 and with the NIRCam and MIRI coronagraphs on JWST, which explains the efficiency of the tip-tilt dither method, the limiting factors on the Roman Coronagraph are expected to be set by variations of higher order aberrations. Indeed, on the one hand, tip-tilt errors will be corrected in real time at 20~Hz and low order aberrations up to Zernike mode Z11 at 2~mHz, and on the other hand, the higher order aberrations are only corrected during the observations of the reference star and are left evolving in open loop during the observation of the science target\cite{Bailey2018,Mennesson2022,Krist2023}. Regular updates of the high-order modes corrections are performed  by interleaving reference star visits during science target observation  (DM touch ups), updates that are calculated at the Science Support Center at IPAC to offload the spacecraft flight control system (ground-in-the-loop control scheme, GITL). This setup favors significant redistribution of the speckle pattern in the dark zone at every touch up, owing to the phase and amplitude  aberrations evolution and to the deferred correction calculations of the GITL control\cite{Krist2023}, which will impede the performance of classical post-processing techniques. 
Furthermore, finding reference stars suitable for the wavefront sensing and control and for the PSF subtraction (bright $V>3$ single star, less than 1~mas angular diameter, within a small delta pitch from the science target to minimizing thermal variations...) is expected to be particularly challenging (Wolff et al. 2024, these proceeding). Relaxing some of these criteria would induce further speckle mismatches between the reference star and the science target acquisition, and also reduce the efficiency of classical RDI techniques. 

In this context, observing strategies that anticipate on these high-order modes variations between the reference star and the science target acquisition are of great interest to optimize post-processing performance. 
The generalization of the small grid dither strategy to high order modes is a robust and methodical approach to do so, as it allows to measure the response of the dark zone to pre-calibrated errors while optimizing the time spent on the bright reference star. This method could also relax some  constraints on the reference star selection while maintaining the nominal post-processed performance. 

To investigate this method and optimize its definition (optimal modes to use, amplitude and frequency of the dithers...), we have developed an end-to-end simulation pipeline, CAPyBARA (Altinier et al. 2024, these proceedings), that includes optical propagation through a coronagraph in a space environment, EFC wavefront control, and image post-processing. The pipeline is currently set up to mimic the Roman Coronagraph, which will enable us to simulate an observing sequence that includes the high-order dither strategy, compare its performance with classical RDI observing strategies, and optimize its parameters. 

In addition, we are collaborating with the HiCAT team at STScI (PI R. Soummer) to demonstrate these methods in a representative lab environment. The HiCAT testbed\cite{Soummer2018,Soummer2022} has been designed precisely to address system-level strategies for high-contrast imaging with complex-aperture space telescopes, in particular multi-stage wavefront control architectures. It routinely achieves $6\times 10^{-8}$ raw contrast limits in broadband visible light (9\% bandwidth around $\lambda=680$~nm) within 4 to 11$\lambda/D$ separations, a performance level very comparable to what will be achieved with the Roman Coronagraph. As such, it is an ideal facility to test and demonstrate our observing strategies before a possible implementation on the Roman Coronagraph. We are currently developing our high-order dither strategy using the digital twin of the HiCAT testbed\cite{Moriarty2018} before an upcoming first implementation on the hardware at STScI (Lau et al. 2024, these proceedings).

\subsection{Timeline}
Our goal as part of the ESCAPE project is to develop these methods in time for an implementation on the Roman Coronagraph for an on-sky demonstration. To reach these goals, we are pushing the development of these methods both through numerical simulations (2022-2025) and lab demonstrations (2024-2027), and we plan to implement them on Roman as ``Guest Algorithm'' after launch in 2027 during the 90-day observing phase after the commissioning of the Coronagraph. If successful, we hope to improve the detection limits below $10^{-9}$ with these post-processing methods, giving access to the atmospheric characterization of a larger sample of mature nearby giant planets.

\section{SUMMARY}
In summary, the ESCAPE project aims at developing observing strategies and post-processing methods making the best out of future high-contrast imaging instruments in space, equipped with state-of-the art wavefront control systems. In particular, we investigate how to use the sensors telemetry and high-order mode dithers on the deformable mirror during the reference star observations to improve the post-processing performance over classical methods. To do so, we use existing data from JWST to test the limits of post-processing methods on instruments not equipped with active wavefront control, and we develop both numerical simulations and lab demonstrations representative of the Roman Coronagraph observations. We plan to implement the high-order dither strategy as guest algorithm on Roman, in order to test it and demonstrate its performance in 2027 after launch. If successful, these new methods could improve the expected detection limits and lead to the detection and characterization of additional nearby giant planets. Such a demonstration would also inform the design and definition of the HWO mission, by possibly relaxing some of its stability constraints.






\acknowledgments       
 
This project is funded by the European Union (ERC, ESCAPE, project No 101044152). Views and opinions expressed are however those of the authors only and do not necessarily reflect those of the European Union or the European Research Council Executive Agency. Neither the European Union nor the granting authority can be held responsible for them.   


\include{bibliography}
\bibliographystyle{spiebib} 

\end{document}

%% file: journaux.tex
\newcommand{\aap}{A\&A}

 \newcommand{\aj}{AJ}
 \newcommand{\apj}{ApJ}
 
 \newcommand{\apjl}{ApJL}
 \newcommand{\pasp}{PASP}

 \newcommand{\mnras}{MNRAS}

 \newcommand{\procspie}{SPIE}

\newcommand{\nat}{Nature}

%% file: main.bbl
\begin{thebibliography}{10}

\bibitem{Kopparapu2018}
{Kopparapu}, R.~K., {H{\'e}brard}, E., {Belikov}, R., {Batalha}, N.~M., {Mulders}, G.~D., {Stark}, C., {Teal}, D., {Domagal-Goldman}, S., and {Mandell}, A., ``{Exoplanet Classification and Yield Estimates for Direct Imaging Missions},'' {\em \apj}~{\bf 856},  122 (Apr. 2018).

\bibitem{Bryson2021}
{Bryson}, S., {Kunimoto}, M., {Kopparapu}, R.~K., {Coughlin}, J.~L., {Borucki}, W.~J., {Koch}, D., {Aguirre}, V.~S., {Allen}, C., {Barentsen}, G., {Batalha}, N.~M., {Berger}, T., {Boss}, A., {Buchhave}, L.~A., {Burke}, C.~J., {Caldwell}, D.~A., {Campbell}, J.~R., {Catanzarite}, J., {Chandrasekaran}, H., {Chaplin}, W.~J., {Christiansen}, J.~L., {Christensen-Dalsgaard}, J., {Ciardi}, D.~R., {Clarke}, B.~D., {Cochran}, W.~D., {Dotson}, J.~L., {Doyle}, L.~R., {Duarte}, E.~S., {Dunham}, E.~W., {Dupree}, A.~K., {Endl}, M., {Fanson}, J.~L., {Ford}, E.~B., {Fujieh}, M., {Gautier}, Thomas~N., I., {Geary}, J.~C., {Gilliland}, R.~L., {Girouard}, F.~R., {Gould}, A., {Haas}, M.~R., {Henze}, C.~E., {Holman}, M.~J., {Howard}, A.~W., {Howell}, S.~B., {Huber}, D., {Hunter}, R.~C., {Jenkins}, J.~M., {Kjeldsen}, H., {Kolodziejczak}, J., {Larson}, K., {Latham}, D.~W., {Li}, J., {Mathur}, S., {Meibom}, S., {Middour}, C., {Morris}, R.~L., {Morton}, T.~D., {Mullally}, F., {Mullally}, S.~E., {Pletcher}, D., {Prsa}, A., {Quinn},
  S.~N., {Quintana}, E.~V., {Ragozzine}, D., {Ramirez}, S.~V., {Sanderfer}, D.~T., {Sasselov}, D., {Seader}, S.~E., {Shabram}, M., {Shporer}, A., {Smith}, J.~C., {Steffen}, J.~H., {Still}, M., {Torres}, G., {Troeltzsch}, J., {Twicken}, J.~D., {Uddin}, A.~K., {Van Cleve}, J.~E., {Voss}, J., {Weiss}, L.~M., {Welsh}, W.~F., {Wohler}, B., and {Zamudio}, K.~A., ``{The Occurrence of Rocky Habitable-zone Planets around Solar-like Stars from Kepler Data},'' {\em \aj}~{\bf 161},  36 (Jan. 2021).

\bibitem{Astro2020}
{National Academies of Sciences, Engineering, and Medicine},  [{\em Pathways to Discovery in Astronomy and Astrophysics for the 2020s}{\nolinebreak\hspace{0.1em}]}, The National Academies Press, Washington, DC (2023).

\bibitem{Traub2010}
{Traub}, W.~A. and {Oppenheimer}, B.~R., ``{Direct Imaging of Exoplanets},'' in [{\em Exoplanets}{\nolinebreak\hspace{0.1em}]},  {Seager}, S., ed.,  111--156 (2010).

\bibitem{Beuzit2019}
{Beuzit}, J.~L., {Vigan}, A., {Mouillet}, D., {Dohlen}, K., {Gratton}, R., {Boccaletti}, A., {Sauvage}, J.~F., {Schmid}, H.~M., {Langlois}, M., {Petit}, C., {Baruffolo}, A., {Feldt}, M., {Milli}, J., {Wahhaj}, Z., {Abe}, L., {Anselmi}, U., {Antichi}, J., {Barette}, R., {Baudrand}, J., {Baudoz}, P., {Bazzon}, A., {Bernardi}, P., {Blanchard}, P., {Brast}, R., {Bruno}, P., {Buey}, T., {Carbillet}, M., {Carle}, M., {Cascone}, E., {Chapron}, F., {Charton}, J., {Chauvin}, G., {Claudi}, R., {Costille}, A., {De Caprio}, V., {de Boer}, J., {Delboulb{\'e}}, A., {Desidera}, S., {Dominik}, C., {Downing}, M., {Dupuis}, O., {Fabron}, C., {Fantinel}, D., {Farisato}, G., {Feautrier}, P., {Fedrigo}, E., {Fusco}, T., {Gigan}, P., {Ginski}, C., {Girard}, J., {Giro}, E., {Gisler}, D., {Gluck}, L., {Gry}, C., {Henning}, T., {Hubin}, N., {Hugot}, E., {Incorvaia}, S., {Jaquet}, M., {Kasper}, M., {Lagadec}, E., {Lagrange}, A.~M., {Le Coroller}, H., {Le Mignant}, D., {Le Ruyet}, B., {Lessio}, G., {Lizon}, J.~L., {Llored}, M.,
  {Lundin}, L., {Madec}, F., {Magnard}, Y., {Marteaud}, M., {Martinez}, P., {Maurel}, D., {M{\'e}nard}, F., {Mesa}, D., {M{\"o}ller-Nilsson}, O., {Moulin}, T., {Moutou}, C., {Orign{\'e}}, A., {Parisot}, J., {Pavlov}, A., {Perret}, D., {Pragt}, J., {Puget}, P., {Rabou}, P., {Ramos}, J., {Reess}, J.~M., {Rigal}, F., {Rochat}, S., {Roelfsema}, R., {Rousset}, G., {Roux}, A., {Saisse}, M., {Salasnich}, B., {Santambrogio}, E., {Scuderi}, S., {Segransan}, D., {Sevin}, A., {Siebenmorgen}, R., {Soenke}, C., {Stadler}, E., {Suarez}, M., {Tiph{\`e}ne}, D., {Turatto}, M., {Udry}, S., {Vakili}, F., {Waters}, L.~B.~F.~M., {Weber}, L., {Wildi}, F., {Zins}, G., and {Zurlo}, A., ``{SPHERE: the exoplanet imager for the Very Large Telescope},'' {\em \aap}~{\bf 631},  A155 (Nov. 2019).

\bibitem{Macintosh2014}
{Macintosh}, B., {Graham}, J.~R., {Ingraham}, P., {Konopacky}, Q., {Marois}, C., {Perrin}, M., {Poyneer}, L., {Bauman}, B., {Barman}, T., {Burrows}, A.~S., {Cardwell}, A., {Chilcote}, J., {De Rosa}, R.~J., {Dillon}, D., {Doyon}, R., {Dunn}, J., {Erikson}, D., {Fitzgerald}, M.~P., {Gavel}, D., {Goodsell}, S., {Hartung}, M., {Hibon}, P., {Kalas}, P., {Larkin}, J., {Maire}, J., {Marchis}, F., {Marley}, M.~S., {McBride}, J., {Millar-Blanchaer}, M., {Morzinski}, K., {Norton}, A., {Oppenheimer}, B.~R., {Palmer}, D., {Patience}, J., {Pueyo}, L., {Rantakyro}, F., {Sadakuni}, N., {Saddlemyer}, L., {Savransky}, D., {Serio}, A., {Soummer}, R., {Sivaramakrishnan}, A., {Song}, I., {Thomas}, S., {Wallace}, J.~K., {Wiktorowicz}, S., and {Wolff}, S., ``{First light of the Gemini Planet Imager},'' {\em Proceedings of the National Academy of Science}~{\bf 111},  12661--12666 (Sept. 2014).

\bibitem{Carter2023}
{Carter}, A.~L., {Hinkley}, S., {Kammerer}, J., {Skemer}, A., {Biller}, B.~A., {Leisenring}, J.~M., {Millar-Blanchaer}, M.~A., {Petrus}, S., {Stone}, J.~M., {Ward-Duong}, K., {Wang}, J.~J., {Girard}, J.~H., {Hines}, D.~C., {Perrin}, M.~D., {Pueyo}, L., {Balmer}, W.~O., {Bonavita}, M., {Bonnefoy}, M., {Chauvin}, G., {Choquet}, E., {Christiaens}, V., {Danielski}, C., {Kennedy}, G.~M., {Matthews}, E.~C., {Miles}, B.~E., {Patapis}, P., {Ray}, S., {Rickman}, E., {Sallum}, S., {Stapelfeldt}, K.~R., {Whiteford}, N., {Zhou}, Y., {Absil}, O., {Boccaletti}, A., {Booth}, M., {Bowler}, B.~P., {Chen}, C.~H., {Currie}, T., {Fortney}, J.~J., {Grady}, C.~A., {Greebaum}, A.~Z., {Henning}, T., {Hoch}, K. K.~W., {Janson}, M., {Kalas}, P., {Kenworthy}, M.~A., {Kervella}, P., {Kraus}, A.~L., {Lagage}, P.-O., {Liu}, M.~C., {Macintosh}, B., {Marino}, S., {Marley}, M.~S., {Marois}, C., {Matthews}, B.~C., {Mawet}, D., {McElwain}, M.~W., {Metchev}, S., {Meyer}, M.~R., {Molliere}, P., {Moran}, S.~E., {Morley}, C.~V., {Mukherjee}, S.,
  {Pantin}, E., {Quirrenbach}, A., {Rebollido}, I., {Ren}, B.~B., {Schneider}, G., {Vasist}, M., {Worthen}, K., {Wyatt}, M.~C., {Briesemeister}, Z.~W., {Bryan}, M.~L., {Calissendorff}, P., {Cantalloube}, F., {Cugno}, G., {De Furio}, M., {Dupuy}, T.~J., {Factor}, S.~M., {Faherty}, J.~K., {Fitzgerald}, M.~P., {Franson}, K., {Gonzales}, E.~C., {Hood}, C.~E., {Howe}, A.~R., {Kuzuhara}, M., {Lagrange}, A.-M., {Lawson}, K., {Lazzoni}, C., {Lew}, B. W.~P., {Liu}, P., {Llop-Sayson}, J., {Lloyd}, J.~P., {Martinez}, R.~A., {Mazoyer}, J., {Palma-Bifani}, P., {Quanz}, S.~P., {Redai}, J.~A., {Samland}, M., {Schlieder}, J.~E., {Tamura}, M., {Tan}, X., {Uyama}, T., {Vigan}, A., {Vos}, J.~M., {Wagner}, K., {Wolff}, S.~G., {Ygouf}, M., {Zhang}, X., {Zhang}, K., and {Zhang}, Z., ``{The JWST Early Release Science Program for Direct Observations of Exoplanetary Systems I: High-contrast Imaging of the Exoplanet HIP 65426 b from 2 to 16 {\ensuremath{\mu}}m},'' {\em \apjl}~{\bf 951},  L20 (July 2023).

\bibitem{Leboulleux2018}
Leboulleux, L., Sauvage, J.-F., Pueyo, L.~A., Fusco, T., Soummer, R., Mazoyer, J., Sivaramakrishnan, A., N'Diaye, M., and Fauvarque, O., ``Pair-based analytical model for segmented telescopes imaging from space for sensitivity analysis,'' {\em Journal of Astronomical Telescopes, Instruments, and Systems}~{\bf 4},  035002 (July 2018).

\bibitem{Laginja2021}
Laginja, I., Soummer, R., Mugnier, L.~M., Pueyo, L., Sauvage, J.-F., Leboulleux, L., Coyle, L., and Knight, J.~S., ``Analytical tolerancing of segmented telescope co-phasing for exo-earth high-contrast imaging,'' {\em Journal of Astronomical Telescopes, Instruments, and Systems}~{\bf 7},  015004 (Jan. 2021).

\bibitem{Pueyo2022}
Pueyo, L., Juanola-Parramon, R., Tumlinson, J., Soummer, R., Laginja, I., Hammel, H.~B., and Mountain, C.~M., ``Coronagraphic detection of earth-like planets with large, actively controlled space telescopes,'' {\em Journal of Astronomical Telescopes, Instruments, and Systems}~{\bf 8},  049002 (Oct. 2022).

\bibitem{Vogt2011}
{Vogt}, F.~P.~A., {Martinache}, F., {Guyon}, O., {Yoshikawa}, T., {Yokochi}, K., {Garrel}, V., and {Matsuo}, T., ``{Coronagraphic Low-Order Wavefront Sensor: Postprocessing Sensitivity Enhancer for High-Performance Coronagraphs},'' {\em \pasp}~{\bf 123},  1434--1441 (Dec. 2011).

\bibitem{Singh2016}
{Singh}, G., {Lozi}, J., {Choquet}, E., {Serabyn}, E., and {Guyon}, O., ``{PSF calibration using the Lyot-based low order wavefront sensor telemetry: first simulations},'' {\em \procspie}~{\bf 9909},  99097K (July 2016).

\bibitem{Soummer2014a}
{Soummer}, R., {Lajoie}, C.-P., {Pueyo}, L., {Hines}, D.~C., {Isaacs}, J.~C., {Nelan}, E.~P., {Clampin}, M., and {Perrin}, M., ``{Small-grid dithering strategy for improved coronagraphic performance with JWST},'' in [{\em Space Telescopes and Instrumentation 2014: Optical, Infrared, and Millimeter Wave}{\nolinebreak\hspace{0.1em}]},  {Oschmann}, Jacobus~M., J., {Clampin}, M., {Fazio}, G.~G., and {MacEwen}, H.~A., eds., {\em Society of Photo-Optical Instrumentation Engineers (SPIE) Conference Series} {\bf 9143},  91433V (Aug. 2014).

\bibitem{Lajoie2016}
{Lajoie}, C.-P., {Soummer}, R., {Pueyo}, L., {Hines}, D.~C., {Nelan}, E.~P., {Perrin}, M., {Clampin}, M., and {Isaacs}, J.~C., ``{Small-grid dithers for the JWST coronagraphs},'' in [{\em Space Telescopes and Instrumentation 2016: Optical, Infrared, and Millimeter Wave}{\nolinebreak\hspace{0.1em}]},  {MacEwen}, H.~A., {Fazio}, G.~G., {Lystrup}, M., {Batalha}, N., {Siegler}, N., and {Tong}, E.~C., eds., {\em Society of Photo-Optical Instrumentation Engineers (SPIE) Conference Series} {\bf 9904},  99045K (July 2016).

\bibitem{Potier2022}
Potier, A., Mazoyer, J., Wahhaj, Z., Baudoz, P., Chauvin, G., Galicher, R., and Ruane, G., ``Increasing the raw contrast of vlt/sphere with the dark hole technique. ii. on-sky wavefront correction and coherent differential imaging,'' {\em \aap}~{\bf 665},  A136 (Sept. 2022).

\bibitem{Stark2019}
{Stark}, C.~C., {Belikov}, R., {Bolcar}, M.~R., {Cady}, E., {Crill}, B.~P., {Ertel}, S., {Groff}, T., {Hildebrandt}, S., {Krist}, J., {Lisman}, P.~D., {Mazoyer}, J., {Mennesson}, B., {Nemati}, B., {Pueyo}, L., {Rauscher}, B.~J., {Riggs}, A.~J., {Ruane}, G., {Shaklan}, S.~B., {Sirbu}, D., {Soummer}, R., {Laurent}, K.~S., and {Zimmerman}, N., ``{ExoEarth yield landscape for future direct imaging space telescopes},'' {\em Journal of Astronomical Telescopes, Instruments, and Systems}~{\bf 5},  024009 (Apr. 2019).

\bibitem{Stark2024}
{Stark}, C.~C., {Mennesson}, B., {Bryson}, S., {Ford}, E.~B., {Robinson}, T.~D., {Belikov}, R., {Bolcar}, M.~R., {Feinberg}, L.~D., {Guyon}, O., {Latouf}, N., {Mandell}, A.~M., {Rauscher}, B.~J., {Sirbu}, D., and {Tuchow}, N.~W., ``{Paths to Robust Exoplanet Science Yield Margin for the Habitable Worlds Observatory},'' {\em arXiv e-prints} ,  arXiv:2405.19418 (May 2024).

\bibitem{LUVOIRTeam2019}
{The LUVOIR Team}, ``{The LUVOIR Mission Concept Study Final Report},'' {\em arXiv e-prints} ,  arXiv:1912.06219 (Dec. 2019).

\bibitem{Gaudi2020}
{Gaudi}, B.~S., {Seager}, S., {Mennesson}, B., {Kiessling}, A., {Warfield}, K., {Cahoy}, K., {Clarke}, J.~T., {Domagal-Goldman}, S., {Feinberg}, L., {Guyon}, O., {Kasdin}, J., {Mawet}, D., {Plavchan}, P., {Robinson}, T., {Rogers}, L., {Scowen}, P., {Somerville}, R., {Stapelfeldt}, K., {Stark}, C., {Stern}, D., {Turnbull}, M., {Amini}, R., {Kuan}, G., {Martin}, S., {Morgan}, R., {Redding}, D., {Stahl}, H.~P., {Webb}, R., {Alvarez-Salazar}, O., {Arnold}, W.~L., {Arya}, M., {Balasubramanian}, B., {Baysinger}, M., {Bell}, R., {Below}, C., {Benson}, J., {Blais}, L., {Booth}, J., {Bourgeois}, R., {Bradford}, C., {Brewer}, A., {Brooks}, T., {Cady}, E., {Caldwell}, M., {Calvet}, R., {Carr}, S., {Chan}, D., {Cormarkovic}, V., {Coste}, K., {Cox}, C., {Danner}, R., {Davis}, J., {Dewell}, L., {Dorsett}, L., {Dunn}, D., {East}, M., {Effinger}, M., {Eng}, R., {Freebury}, G., {Garcia}, J., {Gaskin}, J., {Greene}, S., {Hennessy}, J., {Hilgemann}, E., {Hood}, B., {Holota}, W., {Howe}, S., {Huang}, P., {Hull}, T., {Hunt}, R.,
  {Hurd}, K., {Johnson}, S., {Kissil}, A., {Knight}, B., {Kolenz}, D., {Kraus}, O., {Krist}, J., {Li}, M., {Lisman}, D., {Mandic}, M., {Mann}, J., {Marchen}, L., {Marrese-Reading}, C., {McCready}, J., {McGown}, J., {Missun}, J., {Miyaguchi}, A., {Moore}, B., {Nemati}, B., {Nikzad}, S., {Nissen}, J., {Novicki}, M., {Perrine}, T., {Pineda}, C., {Polanco}, O., {Putnam}, D., {Qureshi}, A., {Richards}, M., {Eldorado Riggs}, A.~J., {Rodgers}, M., {Rud}, M., {Saini}, N., {Scalisi}, D., {Scharf}, D., {Schulz}, K., {Serabyn}, G., {Sigrist}, N., {Sikkia}, G., {Singleton}, A., {Shaklan}, S., {Smith}, S., {Southerd}, B., {Stahl}, M., {Steeves}, J., {Sturges}, B., {Sullivan}, C., {Tang}, H., {Taras}, N., {Tesch}, J., {Therrell}, M., {Tseng}, H., {Valente}, M., {Van Buren}, D., {Villalvazo}, J., {Warwick}, S., {Webb}, D., {Westerhoff}, T., {Wofford}, R., {Wu}, G., {Woo}, J., {Wood}, M., {Ziemer}, J., {Arney}, G., {Anderson}, J., {Ma{\'\i}z-Apell{\'a}niz}, J., {Bartlett}, J., {Belikov}, R., {Bendek}, E., {Cenko}, B.,
  {Douglas}, E., {Dulz}, S., {Evans}, C., {Faramaz}, V., {Feng}, Y.~K., {Ferguson}, H., {Follette}, K., {Ford}, S., {Garc{\'\i}a}, M., {Geha}, M., {Gelino}, D., {G{\"o}tberg}, Y., {Hildebrandt}, S., {Hu}, R., {Jahnke}, K., {Kennedy}, G., {Kreidberg}, L., {Isella}, A., {Lopez}, E., {Marchis}, F., {Macri}, L., {Marley}, M., {Matzko}, W., {Mazoyer}, J., {McCandliss}, S., {Meshkat}, T., {Mordasini}, C., {Morris}, P., {Nielsen}, E., {Newman}, P., {Petigura}, E., {Postman}, M., {Reines}, A., {Roberge}, A., {Roederer}, I., {Ruane}, G., {Schwieterman}, E., {Sirbu}, D., {Spalding}, C., {Teplitz}, H., {Tumlinson}, J., {Turner}, N., {Werk}, J., {Wofford}, A., {Wyatt}, M., {Young}, A., and {Zellem}, R., ``{The Habitable Exoplanet Observatory (HabEx) Mission Concept Study Final Report},'' {\em arXiv e-prints} ,  arXiv:2001.06683 (Jan. 2020).

\bibitem{Bailey2023}
Bailey, V.~P., Bendek, E., Monacelli, B., Baker, C., Bedrosian, G., Cady, E., Douglas, E.~S., Groff, T., Hildebrandt, S.~R., Kasdin, N.~J., Krist, J., Macintosh, B., Mennesson, B., Morrissey, P., Poberezhskiy, I., Subedi, H.~B., Rhodes, J., Roberge, A., Ygouf, M., Zellem, R.~T., Zhao, F., and Zimmerman, N.~T., ``Nancy grace roman space telescope coronagraph instrument overview and status,'' in [{\em Society of Photo-Optical Instrumentation Engineers (SPIE) Conference Series}{\nolinebreak\hspace{0.1em}]},  {\em Society of Photo-Optical Instrumentation Engineers (SPIE) Conference Series} {\bf 12680},  126800T (Oct. 2023).

\bibitem{Roulet2020}
{Roulet}, M., {Hugot}, E., {Atkins}, C., {Marcos}, M., {Lombardo}, S., {Bonnefoi}, A., {Caillat}, A., and {Ferrari}, M., ``{Off-axis parabolas super polished under stress: the case of the Roman Space Telescope coronagraphic instrument mirrors},'' {\em Optics Express}~{\bf 28},  30555 (Oct. 2020).

\bibitem{Riggs2021}
{Riggs}, A.~J.~E., {Bailey}, V., {Moody}, D.~C., {Sidick}, E., {Balasubramanian}, K., {Moore}, D.~M., {Wilson}, D.~W., {Ruane}, G., {Sirbu}, D., {Gersh-Range}, J., {Trauger}, J., {Mennesson}, B., {Siegler}, N., {Bendek}, E., {Groff}, T.~D., {Zimmerman}, N.~T., {Debes}, J., {Basinger}, S.~A., and {Kasdin}, N.~J., ``{Flight mask designs of the Roman Space Telescope coronagraph instrument},'' in [{\em Techniques and Instrumentation for Detection of Exoplanets X}{\nolinebreak\hspace{0.1em}]},  {Shaklan}, S.~B. and {Ruane}, G.~J., eds., {\em Society of Photo-Optical Instrumentation Engineers (SPIE) Conference Series} {\bf 11823},  118231Y (Sept. 2021).

\bibitem{Shi2018}
{Shi}, F., {Seo}, B.-J., {Cady}, E., {Kern}, B., {Lam}, R., {Marx}, D., {Patterson}, K., {Mejia Prada}, C., {Shaw}, J., {Shelton}, C., {Shields}, J., {Tang}, H., and {Truong}, T., ``{WFIRST low order wavefront sensing and control dynamic testbed performance under the flight like photon flux},'' in [{\em Space Telescopes and Instrumentation 2018: Optical, Infrared, and Millimeter Wave}{\nolinebreak\hspace{0.1em}]},  {Lystrup}, M., {MacEwen}, H.~A., {Fazio}, G.~G., {Batalha}, N., {Siegler}, N., and {Tong}, E.~C., eds., {\em Society of Photo-Optical Instrumentation Engineers (SPIE) Conference Series} {\bf 10698},  106982O (July 2018).

\bibitem{Giveon2007a}
{Give'on}, A., {Belikov}, R., {Shaklan}, S., and {Kasdin}, J., ``{Closed loop, DM diversity-based, wavefront correction algorithm for high contrast imaging systems},'' {\em Optics Express}~{\bf 15},  12338 (Jan. 2007).

\bibitem{Giveon2007b}
{Give'on}, A., {Kern}, B., {Shaklan}, S., {Moody}, D.~C., and {Pueyo}, L., ``{Broadband wavefront correction algorithm for high-contrast imaging systems},'' in [{\em Astronomical Adaptive Optics Systems and Applications III}{\nolinebreak\hspace{0.1em}]},  {Tyson}, R.~K. and {Lloyd-Hart}, M., eds., {\em Society of Photo-Optical Instrumentation Engineers (SPIE) Conference Series} {\bf 6691},  66910A (Sept. 2007).

\bibitem{Krist2023}
Krist, J.~E., Steeves, J.~B., Dube, B.~D., Eldorado~Riggs, A.~J., Kern, B.~D., Marx, D.~S., Cady, E.~J., Zhou, H., Poberezhskiy, I.~Y., Baker, C.~W., McGuire, J.~P., Nemati, B., Kuan, G.~M., Mennesson, B., Trauger, J.~T., Saini, N.~S., and Rafels, S.~H., ``End-to-end numerical modeling of the roman space telescope coronagraph,'' {\em Journal of Astronomical Telescopes, Instruments, and Systems}~{\bf 9},  045002 (Oct. 2023).

\bibitem{Bailey2018}
{Bailey}, V.~P., {Bottom}, M., {Cady}, E., {Cantalloube}, F., {de Boer}, J., {Groff}, T., {Krist}, J., {Millar-Blanchaer}, M.~A., {Vigan}, A., {Chilcote}, J., {Choquet}, E., {De Rosa}, R.~J., {Girard}, J.~H., {Guyon}, O., {Kern}, B., {Lagrange}, A.-M., {Macintosh}, B., {Males}, J.~R., {Marois}, C., {Meshkat}, T., {Milli}, J., {N'Diaye}, M., {Ngo}, H., {Nielsen}, E.~L., {Rhodes}, J., {Ruane}, G., {van Holstein}, R.~G., {Wang}, J.~J., and {Xuan}, W., ``{Lessons for WFIRST CGI from ground-based high-contrast systems},'' in [{\em Space Telescopes and Instrumentation 2018: Optical, Infrared, and Millimeter Wave}{\nolinebreak\hspace{0.1em}]},  {Lystrup}, M., {MacEwen}, H.~A., {Fazio}, G.~G., {Batalha}, N., {Siegler}, N., and {Tong}, E.~C., eds., {\em Society of Photo-Optical Instrumentation Engineers (SPIE) Conference Series} {\bf 10698},  106986P (Aug. 2018).

\bibitem{Smith1984}
{Smith}, B.~A. and {Terrile}, R.~J., ``{A circumstellar disk around Beta Pictoris},'' {\em Science}~{\bf 226},  1421--1424 (Dec. 1984).

\bibitem{Schneider1999}
{Schneider}, G., {Smith}, B.~A., {Becklin}, E.~E., {Koerner}, D.~W., {Meier}, R., {Hines}, D.~C., {Lowrance}, P.~J., {Terrile}, R.~J., {Thompson}, R.~I., and {Rieke}, M., ``{NICMOS Imaging of the HR 4796A Circumstellar Disk},'' {\em \apjl}~{\bf 513},  L127--L130 (Mar. 1999).

\bibitem{Marois2006}
{Marois}, C., {Lafreni{\`e}re}, D., {Doyon}, R., {Macintosh}, B., and {Nadeau}, D., ``{Angular Differential Imaging: A Powerful High-Contrast Imaging Technique},'' {\em \apj}~{\bf 641},  556--564 (Apr. 2006).

\bibitem{Sparks2002}
{Sparks}, W.~B. and {Ford}, H.~C., ``{Imaging Spectroscopy for Extrasolar Planet Detection},'' {\em \apj}~{\bf 578},  543--564 (Oct. 2002).

\bibitem{Lafreniere2007}
{Lafreni{\`e}re}, D., {Marois}, C., {Doyon}, R., {Nadeau}, D., and {Artigau}, {\'E}., ``{A New Algorithm for Point-Spread Function Subtraction in High-Contrast Imaging: A Demonstration with Angular Differential Imaging},'' {\em \apj}~{\bf 660},  770--780 (May 2007).

\bibitem{Racine1999}
{Racine}, R., {Walker}, G. A.~H., {Nadeau}, D., {Doyon}, R., and {Marois}, C., ``{Speckle Noise and the Detection of Faint Companions},'' {\em \pasp}~{\bf 111},  587--594 (May 1999).

\bibitem{Soummer2007}
Soummer, R., Ferrari, A., Aime, C., and Jolissaint, L., ``Speckle noise and dynamic range in coronagraphic images,'' {\em \apj}~{\bf 669},  642--656 (Nov. 2007).

\bibitem{Hinkley2007}
{Hinkley}, S., {Oppenheimer}, B.~R., {Soummer}, R., {Sivaramakrishnan}, A., {Roberts}, Lewis~C., J., {Kuhn}, J., {Makidon}, R.~B., {Perrin}, M.~D., {Lloyd}, J.~P., {Kratter}, K., and {Brenner}, D., ``{Temporal Evolution of Coronagraphic Dynamic Range and Constraints on Companions to Vega},'' {\em \apj}~{\bf 654},  633--640 (Jan. 2007).

\bibitem{Flasseur2018}
{Flasseur}, O., {Denis}, L., {Thi{\'e}baut}, {\'E}., and {Langlois}, M., ``{Exoplanet detection in angular differential imaging by statistical learning of the nonstationary patch covariances. The PACO algorithm},'' {\em \aap}~{\bf 618},  A138 (Oct. 2018).

\bibitem{Sing2016}
{Sing}, D.~K., {Fortney}, J.~J., {Nikolov}, N., {Wakeford}, H.~R., {Kataria}, T., {Evans}, T.~M., {Aigrain}, S., {Ballester}, G.~E., {Burrows}, A.~S., {Deming}, D., {D{\'e}sert}, J.-M., {Gibson}, N.~P., {Henry}, G.~W., {Huitson}, C.~M., {Knutson}, H.~A., {Lecavelier Des Etangs}, A., {Pont}, F., {Showman}, A.~P., {Vidal-Madjar}, A., {Williamson}, M.~H., and {Wilson}, P.~A., ``{A continuum from clear to cloudy hot-Jupiter exoplanets without primordial water depletion},'' {\em \nat}~{\bf 529},  59--62 (Jan. 2016).

\bibitem{Lowrance2005}
{Lowrance}, P.~J., {Becklin}, E.~E., {Schneider}, G., {Kirkpatrick}, J.~D., {Weinberger}, A.~J., {Zuckerman}, B., {Dumas}, C., {Beuzit}, J.-L., {Plait}, P., {Malumuth}, E., {Heap}, S., {Terrile}, R.~J., and {Hines}, D.~C., ``{An Infrared Coronagraphic Survey for Substellar Companions},'' {\em \aj}~{\bf 130},  1845--1861 (Oct. 2005).

\bibitem{Lafreniere2009}
{Lafreni{\`e}re}, D., {Marois}, C., {Doyon}, R., and {Barman}, T., ``{HST/NICMOS Detection of HR 8799 b in 1998},'' {\em \apjl}~{\bf 694},  L148--L152 (Apr. 2009).

\bibitem{Soummer2011}
{Soummer}, R., {Hagan}, J.~B., {Pueyo}, L., {Thormann}, A., {Rajan}, A., and {Marois}, C., ``{Orbital Motion of HR 8799 b, c, d Using Hubble Space Telescope Data from 1998: Constraints on Inclination, Eccentricity, and Stability},'' {\em \apj}~{\bf 741},  55 (Nov. 2011).

\bibitem{Hagan2018}
{Hagan}, J.~B., {Choquet}, {\'E}., {Soummer}, R., and {Vigan}, A., ``{ALICE Data Release: A Revaluation of HST-NICMOS Coronagraphic Images},'' {\em \aj}~{\bf 155},  179 (Apr. 2018).

\bibitem{Ren2017}
{Ren}, B., {Pueyo}, L., {Perrin}, M.~D., {Debes}, J.~H., and {Choquet}, {\'E}., ``{Post-processing of the HST STIS coronagraphic observations},'' in [{\em \procspie}{\nolinebreak\hspace{0.1em}]},  {\em \procspie} {\bf 10400},  1040021 (Sept. 2017).

\bibitem{Soummer2012}
{Soummer}, R., {Pueyo}, L., and {Larkin}, J., ``{Detection and Characterization of Exoplanets and Disks Using Projections on Karhunen-Lo{\`e}ve Eigenimages},'' {\em \apjl}~{\bf 755},  L28 (Aug. 2012).

\bibitem{Soummer2014}
{Soummer}, R., {Perrin}, M.~D., {Pueyo}, L., {Choquet}, {\'E}., {Chen}, C., {Golimowski}, D.~A., {Hagan}, J.~B., {Mittal}, T., {Moerchen}, M., {N'Diaye}, M., {Rajan}, A., {Wolff}, S., {Debes}, J., {Hines}, D.~C., and {Schneider}, G., ``{Five Debris Disks Newly Revealed in Scattered Light from the Hubble Space Telescope NICMOS Archive},'' {\em \apjl}~{\bf 786},  L23 (May 2014).

\bibitem{Choquet2016}
{Choquet}, {\'E}., {Perrin}, M.~D., {Chen}, C.~H., {Soummer}, R., {Pueyo}, L., {Hagan}, J.~B., {Gofas-Salas}, E., {Rajan}, A., {Golimowski}, D.~A., {Hines}, D.~C., {Schneider}, G., {Mazoyer}, J., {Augereau}, J.-C., {Debes}, J., {Stark}, C.~C., {Wolff}, S., {N'Diaye}, M., and {Hsiao}, K., ``{First Images of Debris Disks around TWA 7, TWA 25, HD 35650, and HD 377},'' {\em \apjl}~{\bf 817},  L2 (Jan. 2016).

\bibitem{Choquet2017}
{Choquet}, {\'E}., {Milli}, J., {Wahhaj}, Z., {Soummer}, R., {Roberge}, A., {Augereau}, J.-C., {Booth}, M., {Absil}, O., {Boccaletti}, A., {Chen}, C.~H., {Debes}, J.~H., {del Burgo}, C., {Dent}, W.~R.~F., {Ertel}, S., {Girard}, J.~H., {Gofas-Salas}, E., {Golimowski}, D.~A., {G{\'o}mez Gonz{\'a}lez}, C.~A., {Hagan}, J.~B., {Hibon}, P., {Hines}, D.~C., {Kennedy}, G.~M., {Lagrange}, A.-M., {Matr{\`a}}, L., {Mawet}, D., {Mouillet}, D., {N'Diaye}, M., {Perrin}, M.~D., {Pinte}, C., {Pueyo}, L., {Rajan}, A., {Schneider}, G., {Wolff}, S., and {Wyatt}, M., ``{First Scattered-light Images of the Gas-rich Debris Disk around 49 Ceti},'' {\em \apjl}~{\bf 834},  L12 (Jan. 2017).

\bibitem{Choquet2018}
{Choquet}, {\'E}., {Bryden}, G., {Perrin}, M.~D., {Soummer}, R., {Augereau}, J.-C., {Chen}, C.~H., {Debes}, J.~H., {Gofas-Salas}, E., {Hagan}, J.~B., {Hines}, D.~C., {Mawet}, D., {Morales}, F., {Pueyo}, L., {Rajan}, A., {Ren}, B., {Schneider}, G., {Stark}, C.~C., and {Wolff}, S., ``{HD 104860 and HD 192758: Two Debris Disks Newly Imaged in Scattered Light with the Hubble Space Telescope},'' {\em \apj}~{\bf 854},  53 (Feb. 2018).

\bibitem{Marshall2018}
{Marshall}, J.~P., {Milli}, J., {Choquet}, E., {del Burgo}, C., {Kennedy}, G.~M., {Matr{\`a}}, L., {Ertel}, S., and {Boccaletti}, A., ``{Comprehensive Analysis of HD 105, A Young Solar System Analog},'' {\em \apj}~{\bf 869},  10 (Dec. 2018).

\bibitem{Marshall2023}
{Marshall}, J.~P., {Milli}, J., {Choquet}, E., del Burgo, C., Kennedy, G.~M., Kemper, F., Wyatt, M.~C., Kral, Q., and Soummer, R., ``Stirred but not shaken: a multiwavelength view of hd 16743's debris disc,'' {\em \mnras}~{\bf 521},  5940--5951 (June 2023).

\bibitem{Rajan2015}
{Rajan}, A., {Barman}, T., {Soummer}, R., {Brendan Hagan}, J., {Patience}, J., {Pueyo}, L., {Choquet}, {\'E}., {Konopacky}, Q., {Macintosh}, B., and {Marois}, C., ``{Characterizing the Atmospheres of the HR8799 Planets with HST/WFC3},'' {\em \apjl}~{\bf 809},  L33 (Aug. 2015).

\bibitem{Debes2019}
{Debes}, J.~H., {Ren}, B., and {Schneider}, G., ``{Pushing the limits of the coronagraphic occulters on Hubble Space Telescope/Space Telescope Imaging Spectrograph},'' {\em Journal of Astronomical Telescopes, Instruments, and Systems}~{\bf 5},  035003 (July 2019).

\bibitem{Beichman2010}
{Beichman}, C.~A., {Krist}, J., {Trauger}, J.~T., {Greene}, T., {Oppenheimer}, B., {Sivaramakrishnan}, A., {Doyon}, R., {Boccaletti}, A., {Barman}, T.~S., and {Rieke}, M., ``{Imaging Young Giant Planets From Ground and Space},'' {\em \pasp}~{\bf 122},  162 (Feb. 2010).

\bibitem{Girard2020}
{Girard}, J.~H., {Bogat}, E., {Gonzalez-Quiles}, J., {Hildebrandt}, S.~R., {Kane}, S.~R., {Li}, Z., {Turnbull}, M.~C., {Stark}, C., {Mandell}, A., {Meshkat}, T., and {Zimmerman}, N.~T., ``{The Roman exoplanet imaging data challenge: a major community engagement effort},'' in [{\em Space Telescopes and Instrumentation 2020: Optical, Infrared, and Millimeter Wave}{\nolinebreak\hspace{0.1em}]},  {Lystrup}, M. and {Perrin}, M.~D., eds., {\em Society of Photo-Optical Instrumentation Engineers (SPIE) Conference Series} {\bf 11443},  1144337 (Dec. 2020).

\bibitem{Girard2022}
{Girard}, J.~H., {Leisenring}, J., {Kammerer}, J., {Gennaro}, M., {Rieke}, M., {Stansberry}, J., {Rest}, A., {Egami}, E., {Sunnquist}, B., {Boyer}, M., {Canipe}, A., {Correnti}, M., {Hilbert}, B., {Perrin}, M.~D., {Pueyo}, L., {Soummer}, R., {Allen}, M., {Bushouse}, H., {Aguilar}, J., {Brooks}, B., {Coe}, D., {DiFelice}, A., {Golimowski}, D., {Hartig}, G., {Hines}, D.~C., {Koekemoer}, A., {Nickson}, B., {Nikolov}, N., {Kozhurina-Platais}, V., {Pirzkal}, N., {Robberto}, M., {Sivaramakrishnan}, A., {Sohn}, S.~T., {Telfer}, R., {Wu}, C.~R., {Beatty}, T., {Florian}, M., {Hainline}, K., {Kelly}, D., {Misselt}, K., {Schlawin}, E., {Sun}, F., {Williams}, C., {Willmer}, C., {Stark}, C., {Ygouf}, M., {Carter}, A., {Beichman}, C., {Greene}, T.~P., {Roellig}, T., {Krist}, J., {Adams Redai}, J., {Wang}, J., {Clark}, C.~R., {Lewis}, D., and {Ferry}, M., ``{JWST/NIRCam coronagraphy: commissioning and first on-sky results},'' in [{\em Space Telescopes and Instrumentation 2022: Optical, Infrared, and Millimeter
  Wave}{\nolinebreak\hspace{0.1em}]},  {Coyle}, L.~E., {Matsuura}, S., and {Perrin}, M.~D., eds., {\em Society of Photo-Optical Instrumentation Engineers (SPIE) Conference Series} {\bf 12180},  121803Q (Aug. 2022).

\bibitem{Franson2024}
{Franson}, K., {Balmer}, W.~O., {Bowler}, B.~P., {Pueyo}, L., {Zhou}, Y., {Rickman}, E., {Zhang}, Z., {Mukherjee}, S., {Pearce}, T.~D., {Bardalez Gagliuffi}, D.~C., {Biddle}, L.~I., {Brandt}, T.~D., {Bowens-Rubin}, R., {Crepp}, J.~R., {Davidson}, James~W., J., {Faherty}, J., {Ginski}, C., {Horch}, E.~P., {Morgan}, M., {Morley}, C.~V., {Perrin}, M.~D., {Sanghi}, A., {Salama}, M., {Theissen}, C.~A., {Tran}, Q.~H., and {Wolf}, T.~N., ``{JWST/NIRCam 4-5 $\mu$m Imaging of the Giant Planet AF Lep b},'' {\em arXiv e-prints} ,  arXiv:2406.09528 (June 2024).

\bibitem{Perrin2016a}
{Perrin}, M.~D., {Acton}, D.~S., {Lajoie}, C.-P., {Knight}, J.~S., {Lallo}, M.~D., {Allen}, M., {Baggett}, W., {Barker}, E., {Comeau}, T., {Coppock}, E., {Dean}, B.~H., {Hartig}, G., {Hayden}, W.~L., {Jordan}, M., {Jurling}, A., {Kulp}, T., {Long}, J., {McElwain}, M.~W., {Meza}, L., {Nelan}, E.~P., {Soummer}, R., {Stansberry}, J., {Stark}, C., {Telfer}, R., {Welsh}, A.~L., {Zielinski}, T.~P., and {Zimmerman}, N.~T., ``{Preparing for JWST wavefront sensing and control operations},'' in [{\em Space Telescopes and Instrumentation 2016: Optical, Infrared, and Millimeter Wave}{\nolinebreak\hspace{0.1em}]},  {MacEwen}, H.~A., {Fazio}, G.~G., {Lystrup}, M., {Batalha}, N., {Siegler}, N., and {Tong}, E.~C., eds., {\em Society of Photo-Optical Instrumentation Engineers (SPIE) Conference Series} {\bf 9904},  99040F (July 2016).

\bibitem{Rigby2023}
Rigby, J., Perrin, M., McElwain, M., Kimble, R., Friedman, S., Lallo, M., Doyon, R., Feinberg, L., Ferruit, P., Glasse, A., Rieke, M., Rieke, G., Wright, G., Willott, C., Colon, K., Milam, S., Neff, S., Stark, C., Valenti, J., Abell, J., Abney, F., Abul-Huda, Y., Acton, D.~S., Adams, E., Adler, D., Aguilar, J., Ahmed, N., Albert, L., Alberts, S., Aldridge, D., Allen, M., Altenburg, M., {\'A}lvarez-M{\'a}rquez, J., Alves~de Oliveira, C., Andersen, G., Anderson, H., Anderson, S., Argyriou, I., Armstrong, A., Arribas, S., Artigau, E., Arvai, A., Atkinson, C., Bacon, G., Bair, T., Banks, K., Barrientes, J., Barringer, B., Bartosik, P., Bast, W., Baudoz, P., Beatty, T., Bechtold, K., Beck, T., Bergeron, E., Bergkoetter, M., Bhatawdekar, R., Birkmann, S., Blazek, R., Blome, C., Boccaletti, A., B{\"o}ker, T., Boia, J., Bonaventura, N., Bond, N., Bosley, K., Boucarut, R., Bourque, M., Bouwman, J., Bower, G., Bowers, C., Boyer, M., Bradley, L., Brady, G., Braun, H., Breda, D., Bresnahan, P., Bright, S., Britt, C.,
  Bromenschenkel, A., Brooks, B., Brooks, K., Brown, B., Brown, M., Brown, P., Bunker, A., Burger, M., Bushouse, H., Cale, S., Cameron, A., Cameron, P., Canipe, A., Caplinger, J., Caputo, F., Cara, M., Carey, L., Carniani, S., Carrasquilla, M., Carruthers, M., Case, M., Catherine, R., Chance, D., Chapman, G., Charlot, S., Charlow, B., Chayer, P., Chen, B., Cherinka, B., Chichester, S., Chilton, Z., Chonis, T., Clampin, M., Clark, C., Clark, K., Coe, D., Coleman, B., Comber, B., Comeau, T., Connolly, D., Cooper, J., Cooper, R., Coppock, E., Correnti, M., Cossou, C., Coulais, A., Coyle, L., Cracraft, M., Curti, M., Cuturic, S., Davis, K., Davis, M., Dean, B., DeLisa, A., deMeester, W., Dencheva, N., Dencheva, N., DePasquale, J., Deschenes, J., Hunor~Detre, {\"O}., Diaz, R., Dicken, D., DiFelice, A., Dillman, M., Dixon, W., Doggett, J., Donaldson, T., Douglas, R., DuPrie, K., Dupuis, J., Durning, J., Easmin, N., Eck, W., Edeani, C., Egami, E., Ehrenwinkler, R., Eisenhamer, J., Eisenhower, M., Elie, M., Elliott,
  J., Elliott, K., Ellis, T., Engesser, M., Espinoza, N., Etienne, O., Etxaluze, M., Falini, P., Feeney, M., Ferry, M., Filippazzo, J., Fincham, B., Fix, M., Flagey, N., Florian, M., Flynn, J., Fontanella, E., Ford, T., Forshay, P., Fox, O., Franz, D., Fu, H., Fullerton, A., Galkin, S., Galyer, A., Garc{\'\i}a~Mar{\'\i}n, M., Gardner, J.~P., Gardner, L., Garland, D., Garrett, B., Gasman, D., Gaspar, A., Gaudreau, D., Gauthier, P., Geers, V., Geithner, P., Gennaro, M., Giardino, G., Girard, J., Giuliano, M., Glassmire, K., Glauser, A., Glazer, S., Godfrey, J., Golimowski, D., Gollnitz, D., Gong, F., Gonzaga, S., Gordon, M., Gordon, K., Goudfrooij, P., Greene, T., Greenhouse, M., Grimaldi, S., Groebner, A., Grundy, T., Guillard, P., Gutman, I., Ha, K.~Q., Haderlein, P., Hagedorn, A., Hainline, K., Haley, C., Hami, M., Hamilton, F., Hammel, H., Hansen, C., Harkins, T., Harr, M., Hart, J., Hart, Q., Hartig, G., Hashimoto, R., Haskins, S., Hathaway, W., Havey, K., Hayden, B., Hecht, K., Heller-Boyer, C.,
  Henriques, C., Henry, A., Hermann, K., Hernandez, S., Hesman, B., Hicks, B., Hilbert, B., Hines, D., Hoffman, M., Holfeltz, S., Holler, B.~J., Hoppa, J., Hott, K., Howard, J.~M., Howard, R., Hunter, A., Hunter, D., Hurst, B., Husemann, B., Hustak, L., Ilinca~Ignat, L., Illingworth, G., Irish, S., Jackson, W., Jahromi, A., Jakobsen, P., James, L., James, B., Januszewski, W., Jenkins, A., Jirdeh, H., Johnson, P., Johnson, T., Jones, V., Jones, R., Jones, D., Jones, O., Jordan, I., Jordan, M., Jurczyk, S., Jurling, A., Kaleida, C., Kalmanson, P., Kammerer, J., Kang, H., Kao, S.-H., Karakla, D., Kavanagh, P., Kelly, D., Kendrew, S., Kennedy, H., Kenny, D., Keski-kuha, R., Keyes, C., Kidwell, R., Kinzel, W., Kirk, J., Kirkpatrick, M., Kirshenblat, D., Klaassen, P., Knapp, B., Knight, J.~S., Knollenberg, P., Koehler, R., Koekemoer, A., Kovacs, A., Kulp, T., Kumari, N., Kyprianou, M., La~Massa, S., Labador, A., Labiano, A., Lagage, P.-O., Lajoie, C.-P., Lallo, M., Lam, M., Lamb, T., Lambros, S., Lampenfield, R.,
  Langston, J., Larson, K., Law, D., Lawrence, J., Lee, D., Leisenring, J., Lepo, K., Leveille, M., Levenson, N., Levine, M., Levy, Z., Lewis, D., Lewis, H., Libralato, M., Lightsey, P., Link, M., Liu, L., Lo, A., Lockwood, A., Logue, R., Long, C., Long, D., Loomis, C., Lopez-Caniego, M., Lorenzo~Alvarez, J., Love-Pruitt, J., Lucy, A., Luetzgendorf, N., Maghami, P., Maiolino, R., Major, M., Malla, S., Malumuth, E., Manjavacas, E., Mannfolk, C., Marrione, A., Marston, A., Martel, A., Maschmann, M., Masci, G., Masciarelli, M., Maszkiewicz, M., Mather, J., McKenzie, K., McLean, B., McMaster, M., Melbourne, K., Mel{\'e}ndez, M., Menzel, M., Merz, K., Meyett, M., Meza, L., Miskey, C., Misselt, K., Moller, C., Morrison, J., Morse, E., Moseley, H., Mosier, G., Mountain, M., Mueckay, J., Mueller, M., Mullally, S., Murphy, J., Murray, K., Murray, C., Mustelier, D., Muzerolle, J., Mycroft, M., Myers, R., Myrick, K., Nanavati, S., Nance, E., Nayak, O., Naylor, B., Nelan, E., Nickson, B., Nielson, A., Nieto-Santisteban,
  M., Nikolov, N., Noriega-Crespo, A., O'Shaughnessy, B., O'Sullivan, B., Ochs, W., Ogle, P., Oleszczuk, B., Olmsted, J., Osborne, S., Ottens, R., Owens, B., Pacifici, C., Pagan, A., Page, J., Park, S., Parrish, K., Patapis, P., Paul, L., Pauly, T., Pavlovsky, C., Pedder, A., Peek, M., Pena-Guerrero, M., Penanen, K., Perez, Y., Perna, M., Perriello, B., Phillips, K., Pietraszkiewicz, M., Pinaud, J.-P., Pirzkal, N., Pitman, J., Piwowar, A., Platais, V., Player, D., Plesha, R., Pollizi, J., Polster, E., Pontoppidan, K., Porterfield, B., Proffitt, C., Pueyo, L., Pulliam, C., Quirt, B., Quispe~Neira, I., Ramos~Alarcon, R., Ramsay, L., Rapp, G., Rapp, R., Rauscher, B., Ravindranath, S., Rawle, T., Regan, M., Reichard, T.~A., Reis, C., Ressler, M.~E., Rest, A., Reynolds, P., Rhue, T., Richon, K., Rickman, E., Ridgaway, M., Ritchie, C., Rix, H.-W., Robberto, M., Robinson, G., Robinson, M., Robinson, O., Rock, F., Rodriguez, D., Rodriguez Del~Pino, B., Roellig, T., Rohrbach, S., Roman, A., Romelfanger, F., Rose, P.,
  Roteliuk, A., Roth, M., Rothwell, B., Rowlands, N., Roy, A., Royer, P., Royle, P., Rui, C., Rumler, P., Runnels, J., Russ, M., Rustamkulov, Z., Ryden, G., Ryer, H., Sabata, M., Sabatke, D., Sabbi, E., Samuelson, B., Sapp, B., Sappington, B., Sargent, B., Sauer, A., Scheithauer, S., Schlawin, E., Schlitz, J., Schmitz, T., Schneider, A., Schreiber, J., Schulze, V., Schwab, R., Scott, J., Sembach, K., Shanahan, C., Shaughnessy, B., Shaw, R., Shawger, N., Shay, C., Sheehan, E., Shen, S., Sherman, A., Shiao, B., Shih, H.-Y., Shivaei, I., Sienkiewicz, M., Sing, D., Sirianni, M., Sivaramakrishnan, A., Skipper, J., Sloan, G.~C., Slocum, C., Slowinski, S., Smith, E., Smith, E., Smith, D., Smith, C., Snyder, G., Soh, W., Sohn, S.~T., Soto, C., Spencer, R., Stallcup, S., Stansberry, J., Starr, C., Starr, E., Stewart, A., Stiavelli, M., Straughn, A., Strickland, D., Stys, J., Summers, F., Sun, F., Sunnquist, B., Swade, D., Swam, M., Swaters, R., Swoish, R., Taylor, J.~M., Taylor, R., Te~Plate, M., Tea, M., Teague, K.,
  Telfer, R., Temim, T., Thatte, D., Thompson, C., Thompson, L., Thomson, S., Tikkanen, T., Tippet, W., Todd, C., Toolan, S., Tran, H., Trejo, E., Truong, J., Tsukamoto, C., Tustain, S., Tyra, H., Ubeda, L., Underwood, K., Uzzo, M., Van~Campen, J., Vandal, T., Vandenbussche, B., Vila, B., Volk, K., Wahlgren, G., Waldman, M., Walker, C., Wander, M., Warfield, C., Warner, G., Wasiak, M., Watkins, M., Weaver, A., Weilert, M., Weiser, N., Weiss, B., Weissman, S., Welty, A., West, G., Wheate, L., Wheatley, E., Wheeler, T., White, R., Whiteaker, K., Whitehouse, P., Whiteleather, J., Whitman, W., Williams, C., Willmer, C., Willoughby, S., Wilson, A., Wirth, G., Wislowski, E., Wolf, E., Wolfe, D., Wolff, S., Workman, B., Wright, R., Wu, C., Wu, R., Wymer, K., Yates, K., Yeager, C., Yeates, J., Yerger, E., Yoon, J., Young, A., Yu, S., Zak, D., Zeidler, P., Zhou, J., Zielinski, T., Zincke, C., and Zonak, S., ``The science performance of jwst as characterized in commissioning,'' {\em \pasp}~{\bf 135},  048001 (Apr.
  2023).

\bibitem{Mennesson2022}
{Mennesson}, B., {Bailey}, V.~P., {Zellem}, R., {Hildebrandt}, S., {Ygouf}, M., {Rhodes}, J., {Zimmerman}, N., {Nemati}, B., {Gonzalez}, G., {Cady}, E., {Kern}, B., {Koch}, T., {Krist}, J., {Heydorff}, K., {Luchik}, T., {Mok}, F., {Morrissey}, P., {Poberezhskiy}, I., {Riggs}, A.~J., {Shi}, F., {Zhao}, F., {Akeson}, R., {Armus}, L., {Greenbaum}, A., {Ingalls}, J., and {Lowrance}, P., ``{The Roman Space Telescope coronagraph technology demonstration: current status and relevance to future missions},'' in [{\em Space Telescopes and Instrumentation 2022: Optical, Infrared, and Millimeter Wave}{\nolinebreak\hspace{0.1em}]},  {Coyle}, L.~E., {Matsuura}, S., and {Perrin}, M.~D., eds., {\em Society of Photo-Optical Instrumentation Engineers (SPIE) Conference Series} {\bf 12180},  121801W (Aug. 2022).

\bibitem{Soummer2018}
{Soummer}, R., {Brady}, G.~R., {Brooks}, K., {Comeau}, T., {Choquet}, {\'E}., {Dillon}, T., {Egron}, S., {Gontrum}, R., {Hagopian}, J., {Laginja}, I., {Leboulleux}, L., {Perrin}, M.~D., {Petrone}, P., {Pueyo}, L., {Mazoyer}, J., {N'Diaye}, M., {Riggs}, A.~J.~E., {Shiri}, R., {Sivaramakrishnan}, A., {St. Laurent}, K., {Valenzuela}, A.-M., and {Zimmerman}, N.~T., ``{High-contrast imager for complex aperture telescopes (HiCAT): 5. first results with segmented-aperture coronagraph and wavefront control},'' in [{\em \procspie}{\nolinebreak\hspace{0.1em}]},  {\em \procspie} {\bf 10698},  106981O (Aug 2018).

\bibitem{Soummer2022}
{Soummer}, R., {Por}, E.~H., {Pourcelot}, R., {Redmond}, S., {Laginja}, I., {Will}, S.~D., {Perrin}, M.~D., {Pueyo}, L., {Sahoo}, A., {Petrone}, P., {Brooks}, K.~J., {Fox}, R., {Klein}, A., {Nickson}, B., {Comeau}, T., {Ferrari}, M., {Gontrum}, R., {Hagopian}, J., {Leboulleux}, L., {Leongomez}, D., {Lugten}, J., {Mugnier}, L.~M., {N'Diaye}, M., {Nguyen}, M., {Noss}, J., {Sauvage}, J.-F., {Scott}, N., {Sivaramakrishnan}, A., {Subedi}, H.~B., and {Weinstock}, S., ``{High-contrast imager for complex aperture telescopes (HiCAT): 8. Dark zone demonstration with simultaneous closed-loop low-order wavefront sensing and control},'' in [{\em Space Telescopes and Instrumentation 2022: Optical, Infrared, and Millimeter Wave}{\nolinebreak\hspace{0.1em}]},  {Coyle}, L.~E., {Matsuura}, S., and {Perrin}, M.~D., eds., {\em Society of Photo-Optical Instrumentation Engineers (SPIE) Conference Series} {\bf 12180},  1218026 (Aug. 2022).

\bibitem{Moriarty2018}
{Moriarty}, C., {Brooks}, K., {Soummer}, R., {Perrin}, M., {Comeau}, T., {Brady}, G., {Gontrum}, R., and {Petrone}, P., ``{High-contrast imager for complex aperture telescopes (HiCAT): 6. software control infrastructure and calibration},'' in [{\em Space Telescopes and Instrumentation 2018: Optical, Infrared, and Millimeter Wave}{\nolinebreak\hspace{0.1em}]},  {Lystrup}, M., {MacEwen}, H.~A., {Fazio}, G.~G., {Batalha}, N., {Siegler}, N., and {Tong}, E.~C., eds., {\em Society of Photo-Optical Instrumentation Engineers (SPIE) Conference Series} {\bf 10698},  1069853 (Aug. 2018).

\end{thebibliography}
